# MACHINE LEARNING POTENTIALS: A ROADMAP TOWARD NEXT-GENERATION BIOMOLECULAR SIMULATIONS

Gianni De Fabritiis[1,2,3]

[1] Computational Science Laboratory, Universitat Pompeu Fabra, Barcelona Biomedical Research Park (PRBB), C Dr. Aiguader 88, 08003 Barcelona, Spain.

[2] Acellera Therapeutics, 38350 Fremont Blvd 203, Fremont CA, 94536 USA.

[3] ICREA, Passeig Lluis Companys 23, 08010 Barcelona, Spain.

Email: g.defabritiis@gmail.com

*Machine learning potentials offer a revolutionary, unifying framework for molecular simulations across scales, from quantum chemistry to coarse-grained models. Here, I explore their potential to dramatically improve accuracy and scalability in simulating complex molecular systems. I discuss key challenges that must be addressed to fully realize their transformative potential in chemical biology and related fields.*

Machine learning (ML) potentials are poised to revolutionize molecular simulations across multiple scales, leveraging the innate ability of neural networks to capture complex correlations in high-dimensional spaces. Here, I discuss how these advanced models can dramatically improve the accuracy and efficiency of simulations, from quantum-mechanical calculations to coarse-grained dynamics. By bridging the gap between atomistic detail and macroscopic behavior, ML potentials promise to unlock new insights into molecular processes, drug discovery, and materials design [Duignan24]. I highlight recent successes, current challenges, and future directions in this rapidly evolving field, emphasizing the transformative potential of ML-driven simulations in chemical biology and related disciplines.

Classical atomistic molecular mechanics potentials are mathematical models used to describe the energy and forces between atoms in a molecular system. Unlike quantum chemistry methods, which explicitly treat electronic structures (Figure 1a), these potentials coarse-grain the effects of electrons into simplified interaction terms between atomic centers (Figure 1b). Typically, they consist of bonded terms (describing bond stretching, angle bending, and torsional rotations) and non-bonded terms (such as van der Waals and electrostatic interactions). This simplification allows for the simulation of much larger systems and longer timescales than quantum chemistry methods, but at the cost of reduced accuracy and the inability to model electronic processes. The parameters for these potentials are usually derived from experimental data or higher-level quantum calculations. While classical potentials have been invaluable in many areas of molecular modeling [Lee09], their fixed functional forms limit their ability to capture complex, many-body interactions accurately across diverse chemical environments.

To simulate even larger macromolecules and their assemblies researchers have developed different coarse-grained functional forms and parameterizations that further reduce computational complexity. Traditional coarse-grained force fields, group multiple heavy atoms into single interaction sites or "beads" using predefined [Marrink07] and functional forms. These models significantly extend the accessible time and length scales of simulations, enabling the study of complex biological processes like protein folding, membrane dynamics, and macromolecular assembly. However, the ad hoc nature of their functional form and parameterization can limit transferability and accuracy.

Neural network potentials offer an exciting, unifying language for molecular simulations across scales. This paradigm shift allows us to view classical molecular mechanics and coarse-grained simulations as instances of the same fundamental learning process, starting from quantum mechanical principles and progressively abstracting to coarser models (Figure 1abc). By framing potential energy surfaces as learnable functions, we can systematically derive models at various levels of granularity while maintaining quantitative agreement with nature in the remaining degrees of freedom. This hierarchical learning approach enables a seamless transition between scales, from electronic structure calculations to atomistic simulations, and further to mesoscale and macroscale models. The flexibility of neural networks in capturing complex, many-body interactions allows for the development of transferable potentials that can adapt to diverse chemical environments. Moreover, this unified framework facilitates the integration of experimental data and theoretical insights at multiple scales, potentially leading to more accurate and physically consistent multi-scale models by preserving the energetics of more accurate (Figure 1d) to coarser models of proteins (Figure 1e) and bead-like C-alpha models (Figure 1f). As a result, machine learning potentials are poised to reshape our conceptual approach to molecular modeling, offering a coherent methodology for bridging quantum mechanics, statistical mechanics, and macroscopic phenomena.

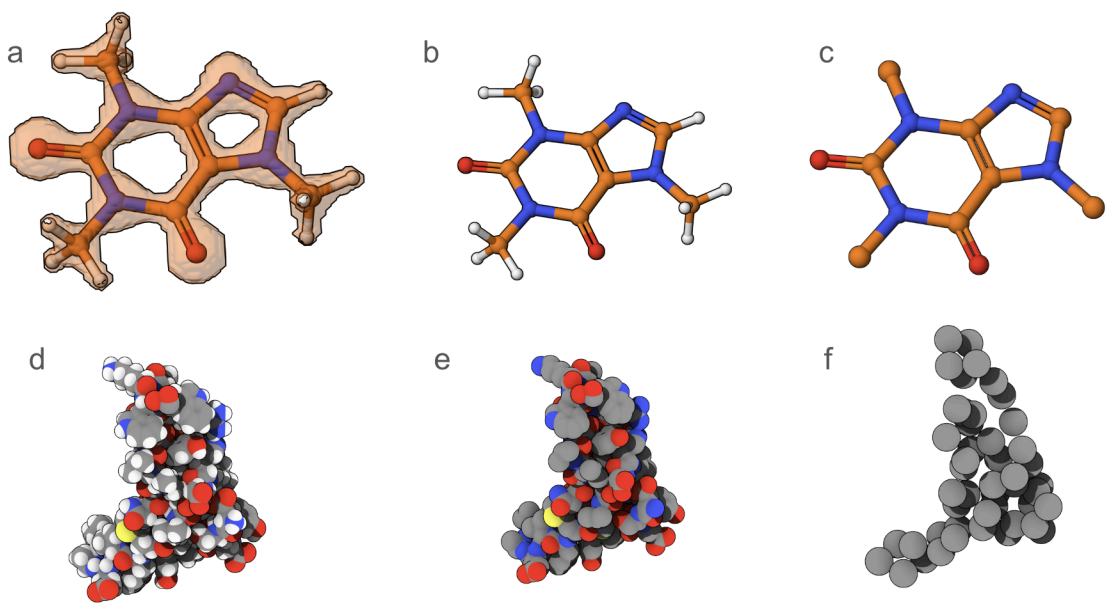

**Figure 1.** Molecular levels of description. In a) a caffeine molecule represented with all atoms and the electronic density which can be resolved by different levels of quantum chemistry approximations. In b) the all-atom representation of molecular mechanics where Newton's equations are used for the dynamics and the electronic degrees of freedom are averaged out. In c) a coarser representation where hydrogen degrees of freedom are eliminated. d) All-atom representation for a small peptide, b) same but hydrogens are implicit, f) a C-alpha representation. All levels can be treated in a unified manner by machine learning potentials just changing the representation and data for training.

**Machine Learning Potentials.** There are two classes of machine learning potentials, parametric methods based on artificial neural networks and nonparametric methods, usually based on kernel methods. Here, I will mainly focus on parametric, neural network potentials (NNPs) because they have the advantage of scaling well with large amounts of data, while kernel methods usually work best in a data scarcity regime. Both types of machine learning potentials can be trained to approximate the energy function of quantum mechanics (QM) at a fraction of the computation cost. The first important contributions to neural network potentials are rooted in the Behler-Parrinello (BP) representation [Behler07] and the seminal work from [Rupp12]. One of the earliest transferable machine learning potentials for biomolecules [Smith17] is based on a modified BP. A second generation of methods, mainly developed in the field of materials science and quantum chemistry, uses graph networks. See [Duval23], for a review of recent methods.

All the methods perform a simple potential fitting by approximating via a machine learning model the high-dimensional potential energy of the system, $V(R) = E$, where $R = (r_1, ..., r_N) \in R^{3N}$, E is the energy and N is the number of atoms in the system. A neural network (NN), given enough parameters, is capable of interpolating any continuous function [Cybenko1989]. Interestingly, the input space can be very highly dimensional, therefore a neural network fitting molecular potential allows to direct representation of many-body interactions. Traditional force fields represent interactions as two-body interactions between two pairs of atoms. Exceptions are angles (3-body) and dihedral (4-body) interactions, but these are quite local interactions between covalently bonded atoms. This is done for computational efficiency, but it has inherent approximations as the quantum wave function contains all the degrees of freedom.

Importantly, the NNP has to respect some physical laws, e.g. translation and rotational invariance, conservation of energy and momentum and invariance upon change of order of atoms. One simple way to obtain conservation of energy is to train the machine learning model by minimizing the mean square error (MSE) between the predicted and real energies, $Loss = |V(R) - NNP(R)|^2$ and then use the fact that forces can be directly computed via a gradient of the NNP, $F_i(R) = -\partial_{R_i} NNP(R)$. In practice, to improve convergence also a force term is included in the loss as the gradient of the energy. The training process can be repeated and automated. For example, the potential can be trained using an active learning algorithm to identify additional molecules and their conformation [Schwalbe-Koda21], which needs to be included in a training set.

More recently, machine learning potentials have emerged for coarser representations. Capable of learning both the interaction potentials and the optimal coarse-grained representations (Figure 1def) directly from data [Wang19b]. These data-driven approaches can automatically identify relevant collective variables and construct accurate, transferable potentials that capture the essential physics of the system across multiple scales. By leveraging large datasets of high-resolution simulations or experimental data, ML-based coarse-graining methods offer a systematic path to developing next-generation models for complex biological systems, potentially bridging the gap between atomistic detail and cellular-scale phenomena. The price to pay by systematically coarse-graining the degrees of freedom of the dynamical system is that the function $V(X) = E$ becomes a stochastic function, with a given probability distribution $p(E|X)$ and it is noisier and harder to train. This is because the system energy depends on degrees of freedom which are no longer available. In recent multiple works, the force-matching method was proposed to learn the potential of mean force over the remaining degrees of freedom. Using this loss [Wang19a] to train an NNP based on α-carbons it was possible to simulate the folding of fast-folding miniproteins [Majewski23]. Most importantly, the NNP trained with the force-matching method successfully reproduced the energetics landscape of the all-atom protein and correctly identified the folded state as a preferred minimum of energy [Majewski23]. More recently, this same approach has been shown to generalize outside of the training set [Charron23].

## Challenges

Despite the promising outlook for neural network potentials, several significant challenges must be addressed to fully realize their potential in molecular simulations. Firstly, the computational cost of evaluating these potentials is substantially higher than that of classical functional-based potentials, due to the orders-of-magnitude difference in the number of parameters (100-1,000 for molecular mechanics versus 100,000-1,000,000 for neural networks). This necessitates the development of faster, more efficient machine learning potentials to make them competitive for large-scale simulations. Secondly, these models require vast amounts of training data. While quantum mechanics provides a fundamental description of nature, generating the billions of data points needed for robust training remains computationally expensive, but possible.

Machine learning potentials have emerged earlier in material sciences because the extended number of elements made it harder to use fixed-form potentials as in protein simulations. However, the challenges for materials are somehow different from the ones in biomolecular simulations. It is useful to discuss the differences. The baseline alternatives to NNPs in material sciences are expensive density functional theory (DFT) calculations, these are a lot slower than biomolecular simulations, the timescales of the simulations are short and the systems are smaller. The "coarse-graining" of the electronic structure into the nuclei degrees of freedom is harder pushing the limit towards designing very accurate machine learning architectures, no matter the cost. On the contrary, biomolecular simulations involve a hundred thousand atoms and a timescale of the order of microseconds to milliseconds of simulated time. A classical molecular mechanics simulations code will perform a single update step on a modern GPU below a millisecond of compute time and will iterate for billions of steps, while DFT calculation can take minutes for a single configuration. It is therefore not surprising that while machine learning potentials for materials and biological simulations are strongly related, the best solution might be different in each case.

**Co-evolution of models, software, and hardware.** The software and hardware ecosystem for machine learning simulations is still in its infancy. Current deep learning frameworks like PyTorch and JAX are optimized for efficient batching rather than the low-latency operations required for molecular dynamics simulations, where potentials must be evaluated billions of times to reach interesting timescales. Achieving millisecond-level elapsed time per step is essential for these methods to compete with established simulation techniques. Overcoming these limitations in speed, data generation, and software infrastructure is paramount to unlocking the full potential of machine learning-driven molecular simulations.

Given the massively higher number of floating-point operations (Flops) required to compute forces in NNPs, it is hard to believe, at first, that these potentials will ever get to the same speed as classical molecular mechanics force fields. However, there are some possible paths. One main advantage of NNPs is that the model architecture is not fixed, it can be adapted. As an example, we always tried to build models [Simeon24] using Cartesian-based equivariance to make sure that all operations were basic matrix multiplications instead of spherical harmonics. In Nvidia GPUs, for example, matrix multiplications using tensor cores have seven times the peak flops of normal CUDA cores where MM calculations are performed. So far, tensor cores arithmetics is based on a reduced floating point format (TF32) compared to FP32 of CUDA cores. It is unclear if the level of noise introduced is excessive for physics. In any case, some level of error control is required. In the early days of accelerators and GPUs without double precision units [DeFabritiis07, Harvey09a], it was common to perform critical operations using two float32, e.g. summing up positive and negative force contributions separately and only doing a final addition between the two numbers once to avoid cancellation errors. Some similar tricks might be required for using tensor cores for neural network potentials. Furthermore, even though the NNP architecture is based on matrix multiplications, it is difficult to fill tensor cores due to other constraints, like memory loads, etc. A coevolution of GPU hardware to satisfy these types of calculations is required to achieve speeds comparable with MM potentials. This is not unheard of, hardware companies routinely run their GPU designs against popular computational codes to optimize the design inside the useful application space, but more work is required here.

Alternatively, a more focused approach may be required. It is interesting to see new machine learning frameworks like TinyGrad [TinyGrad24] which advocate their advantage in depending only on a dozen hardware primitives, compared to hundreds of PyTorch. This approach would facilitate building task-specific NNP-potential hardware by simply providing hardware instructions for those. Similar in spirit to what the Anton chip [Shaw08] is for molecular mechanics potentials, a machine learning potential chip could be created.

The accuracy of current architectures is probably enough for biomolecular simulations and focusing on models which are even more accurate is not productive. Instead, it would be wise to trade some of the accuracy for speed. Furthermore, NNPs usually require a smaller timestep for stability compared to their classical counterparts. Smoother machine learning potentials are needed to get closer to 4 fs timestep as in classical potentials. One key issue is to reduce the memory footprint of some of these NNP architectures if they need to be able to simulate hundreds of thousands of atoms in the future.

**Long-range electrostatics, charges, and spins.** Contrary to simple intuition, having explicit charge representation is not required to produce accurate potentials also for charged molecules [Simeon24a, Kovács23]. Most accurate models currently do not necessarily encode charges or spins. However, there is a problem of degeneracy in the datasets which obliges careful data curation [Simeon24b] if these quantum state variables are not available to the machine learning model. Furthermore, whether explicit treatment of long-range electrostatic is needed is still unclear. On the one hand, we know that in biomolecular systems charge interactions can span over more than 25A, particularly in membrane protein simulations where water screening is not acting across the membrane. There are known ways to deal with the problem by using GPU-efficient long-range Particle-Mesh-Ewald on GPUs [Harvey09b]. This implies the calculations of partial charges which is common in classical molecular simulations. Dynamical algorithms to update the partial charges depending on the conformation are also possible [Unke19] but they are more expensive. On the other hand, partial charges are not physical, in the sense that they are not directly quantum observables and they are often inferred from molecular bonds which are themselves not quantum observables. Machine learning potentials may be capable of handling charge distributions implicitly based on nuclei positions and total charge (and spins), provided that they have a large enough receptive field or enough message-passing layers. Given that essentially all modern molecular dynamics codes run best on a single GPU, this might not be a problem.

**Data generation.** One could argue that unlike any other biological-oriented field relying on experiments, in this case, data is not a problem [Pérez18]. We are lucky enough to know the fundamental equations of motion of atoms and have decades of quantum chemistry approximations at different levels of accuracy. It is not difficult to think that once these models become useful in applications, large datasets of QM data can be generated at different levels of accuracy. It might be long and expensive but not an impossible challenge compared to the amount of calculations that these days are used for large language models. The only remaining question is the scalability of the potential for larger molecules not available in the training set. If the NNP does not generalize well to larger molecules, this could present a real challenge for data accurate generating, with quantum chemistry approximations scaling polynomially with the number of atoms.

Similarly, generating all-atom classical molecular dynamics datasets to train coarse-grained NNPs of proteins should not be impossible. Molecular mechanics atomistic models of proteins are well-tuned and fast enough to produce huge molecular dynamics datasets on demand. In mdCATH [Mirarchi24], most CATH domains have been simulated at different temperatures integrating over 60 ms of simulation data. It is possible to simulate ten times this quantity until the bottleneck becomes training. Finally, more methodological improvements and methods might be able to reach or integrate with the approach, for example, diffusion models applied at force field parameterization [Durumeric24] for data optimization.

## Outlook

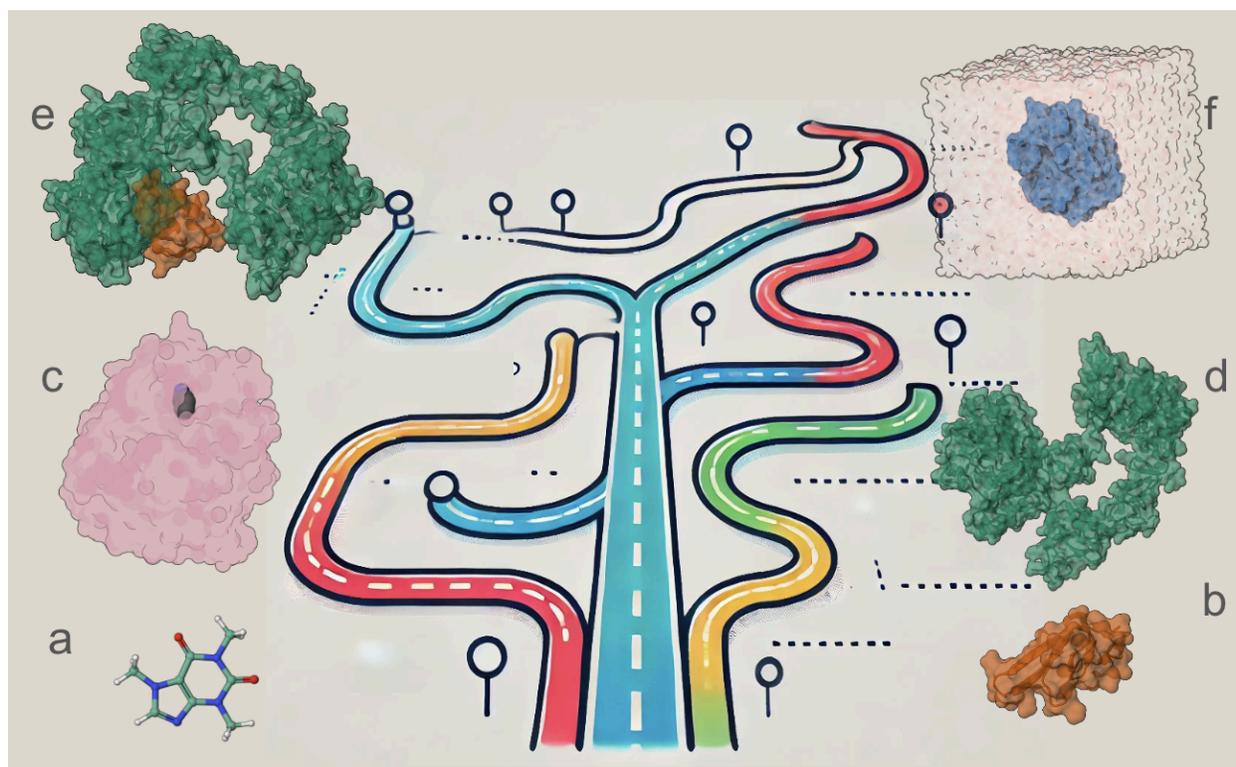

**Figure 2.** A possible roadmap of progressively more complex and useful applications that machine learning potentials should be able to handle, using all-atoms and coarser representations. In a) small molecules, simulation of small molecules is already possible. In b), NNPs can simulate small proteins and peptides at the all-atom level and coarse-grained. c) Complex formation and potency of molecular recognition of protein and small molecules are particularly useful for therapeutics. d) The state and configurations of multi-domain proteins become within reach of coarse-grained NNPs. e) Enable the efficient simulation of protein-protein complexes. f) All parts of the systems are described by very accurate NNPs trained on QM data.

From a technical standpoint, speed remains the utmost concern in biomolecular simulations based on neural network potentials. While artificial neural network potentials are good high-dimensional function approximators, they might be an overkill. Learned potentials from symbolic regressions [Cranmer20], or spline-based methods could produce similar accuracies at much less costs. The co-evolution of models and hardware is likely key if NNP succeeds, possibly propelled by adjacent applications in machine learning.

As mentioned earlier, it is unlikely that data will be a limiting factor since we can generate it using physics-based simulations. Although this process might be costly, it is not prohibitive and, historically, applying machine learning methods to large datasets has proven to be successful.

The software infrastructure for running these new generation models is still in its early stages, with several options available, though none may be ideal. At this point, investing time in extensive optimizations seems premature, as improved models continue to emerge each year. However, due to the simplicity of the approach, developing a new optimized molecular simulation engine for machine learning potentials is feasible, even from scratch, by utilizing components from legacy codes. The most challenging aspect remains the optimization of the neural network potential itself, which is heavily dependent on the specific model.

In Figure 2, I propose potential key milestones and timelines. Machine learning potential-enabled biomolecular simulations will span multiple scales, from atomistic and molecular descriptions to protein and macromolecular assemblies. This advancement will enable simulations of currently unreachable systems with progressively increasing accuracy.

Unsurprisingly, neural network potentials (NNPs) are most advanced and useful for small molecules (Figure 2a). The smaller size of these molecules makes data generation easier, and the models are accurate enough to match quantum mechanics calculations at the same level of accuracy [Kovács23]. Incorporating reactivity [Yang24] could expand their applications, such as in synthesis.

Classical molecular mechanics force fields are excellent for protein simulations. However, it would be beneficial to simulate all solutes using NNPs while describing the solvent with classical potentials (Figure 2b). While simulations of proteins with NNPs have not yet reached the level of classical potentials, we have successfully simulated small proteins and peptides using coarse-grained models.

The first significant application of NNPs in chemical biology will likely be in determining potency [Sabanés24]. Using neural network potentials to represent only the drug molecule can achieve state-of-the-art correlations for relative binding free energy calculations (Figure 2c), even when the rest of the system is modeled with molecular mechanics. Since non-bonded interactions are still handled by molecular mechanics, the improvement stems solely from better quantification of ligand strain upon binding. Due to computational costs, the NNP/MM [Galvelis23] scheme is currently the only practical approach for using NNPs in drug discovery. However, it is exciting to consider the future possibilities as more molecular mechanics components are replaced, potentially achieving such accurate calculations that expensive experimental testing will serve merely as validation.

Studying the conformational dynamics of multidomain proteins (Figure 2d) with coarse-grained, machine-learned models will allow the exploration of previously unseen experimental structures and offer computational capabilities that are not yet routinely achievable. I expect that this will become possible for exemplar cases within the next few years if coarse-grained NNPs have to become important for applications.

Using coarse-grained NNPs for protein-protein recognition (Figure 2e) is more challenging due to the limited availability of all-atom data for training [Plattner17]. However, it might be feasible to simulate protein-protein recognition with optimized datasets and efficient training schemes, but there is no clear path to simulating these systems with the current state-of-the-art. Nevertheless, integrating physical terms for electrostatics would aid in recognition, and coarse-grained NNPs might resolve the process at a short range.

Ultimately, the goal is to simulate entire systems at an all-atom level, including the solvent, with the accuracy of quantum chemistry (Figure 2f) using NNPs. However, there are currently no practical methods to compete with the speed of classical potentials on such a big number of atoms, and it is unclear if this would significantly improve accuracy for large systems. I am hopeful that future technical and methodological advancements will make this feasible.

## Disclaimers

Some of these ideas might be proven incorrect but hopefully will help the readers shape the future of NNPs. The field is moving fast and this commentary has limited space, some important contributions might have been missed. During the writing of this manuscript, I have used both Claude3.5 Sonnet and GTP-4o.